\documentclass[%
 aip,
 amsmath,amssymb,
 reprint,%
]{revtex4-1}

\usepackage{graphicx}
\usepackage{dcolumn}
\usepackage{bm}
\usepackage[normalem]{ulem}
\usepackage[utf8]{inputenc}
\usepackage[T1]{fontenc}
\usepackage{amsthm}
\usepackage{mathptmx}
\usepackage{etoolbox}
\usepackage[dvipsnames,table,xcdraw]{xcolor}
\usepackage{algorithm2e}
\usepackage{dsfont}

\newcommand{\ppi}{{\tilde{\pi}}}
\newcommand{\hh}{{\tilde{h}}}
\newcommand{\vv}{{\tilde{v}}}
\newcommand{\qq}{{\tilde{q}}}

\newcommand{\TT}{{\tilde{T}}}

\newcommand{\appropto}{\mathrel{\vcenter{
  \offinterlineskip\halign{\hfil$##$\cr
    \propto\cr\noalign{\kern2pt}\sim\cr\noalign{\kern-2pt}}}}}
    
\newcommand{\mfpt}{\mathrm{MFPT}}

\makeatletter
\def\@email#1#2{%
 \endgroup
 \patchcmd{\titleblock@produce}
  {\frontmatter@RRAPformat}
  {\frontmatter@RRAPformat{\produce@RRAP{*#1\href{mailto:#2}{#2}}}\frontmatter@RRAPformat}
  {}{}
}%
\makeatother
\begin{document}

\preprint{AIP/123-QED}

\title[Weighted ensemble: Recent mathematical developments]{Weighted ensemble: Recent mathematical developments}

\author{D. Aristoff}
\affiliation{Mathematics, Colorado State University}
\author{J. Copperman}
\affiliation{Biomedical Engineering, Oregon Health Sciences University}%
\author{G. Simpson}
\affiliation{Mathematics, Drexel University}%
\author{R. J. Webber}
\affiliation{Computing \& Mathematical Sciences, California Institute of Technology}
\author{D. M. Zuckerman}
\affiliation{Biomedical Engineering, Oregon Health Sciences University}


\begin{abstract}
The weighted ensemble (WE) method, an enhanced sampling approach based on periodically replicating and pruning trajectories in a set of parallel simulations, has grown increasingly popular for computational biochemistry problems, due in part to improved hardware and the availability of modern software.  Algorithmic and analytical improvements have also played an important role, and progress has accelerated in recent years.  Here, we discuss and elaborate on the WE method from a mathematical perspective, highlighting recent results which have begun to yield greater computational efficiency.
Notable among these innovations are variance reduction approaches that optimize trajectory management for systems of arbitrary dimensionality.
\end{abstract}

\maketitle

\section{Introduction}

Weighted ensemble (WE) \cite{huber1996weighted} is an enhanced sampling method employing  multiple trajectories of a stochastic dynamics  to estimate \emph{mean first passage times} (MFPTs) and related statistics.
WE can be applied to any 
stochastic dynamics model\cite{zhang2010weighted}, such as Langevin dynamics in a molecular system \cite{huber1996weighted} or continuous-time jump dynamics in a reaction network \cite{donovan2013efficient}.
In biomolecular systems,
WE has enabled estimation of MFPTs that are orders of magnitude larger than the combined lengths of the individual WE trajectories. \cite{zhang2007efficient,zwier2011efficient,suarez2014simultaneous,lotz2018unbiased}
WE has also recently been used to elucidate the spike opening dynamics in the SARS CoV-2 virus. \cite{sztain2021glycan} 

\begin{figure}
\includegraphics[width=\columnwidth]{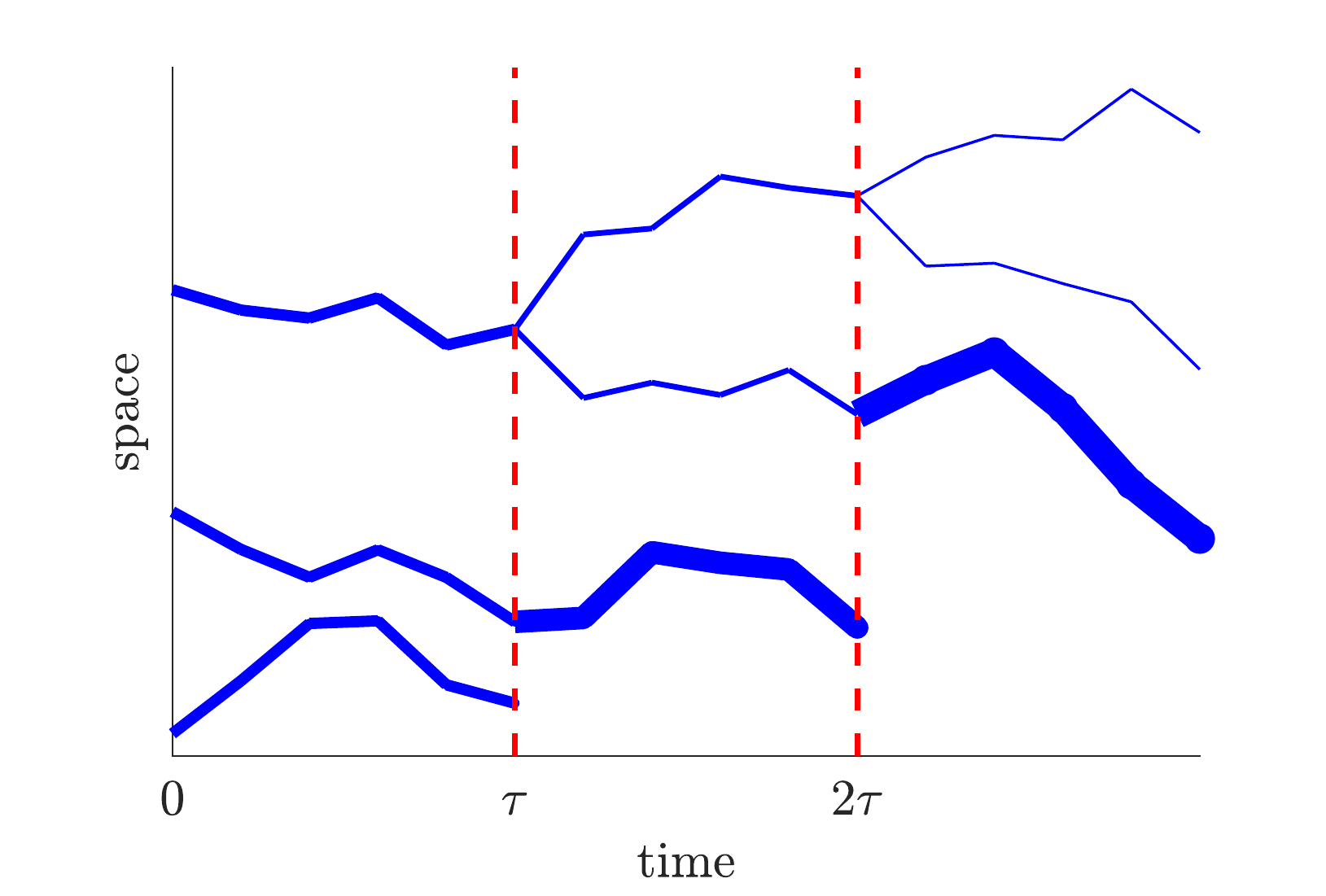}
\caption{\label{fig:splitting_merging} 
{\em An illustration of splitting and merging of 
weighted trajectories.}
Weights are represented by the
widths of trajectories. Two 
splitting and merging steps 
are shown, in which the top 
trajectory is split and 
the bottom trajectories 
are merged. Trajectory weights are combined with merging and divided with splitting.}
\end{figure}

WE is based on \emph{splitting} and \emph{merging}, as shown in Figure~\ref{fig:splitting_merging}.
During splitting, ``favorable'' or ``interesting'' trajectories, according to a user definition, are replicated.  
During merging, the ``less favorable'' or ``redundant'' trajectories are randomly eliminated from the ensemble.
Trajectories are then re-weighted to
preserve the statistics of the path ensemble.\cite{zhang2010weighted} 

The first splitting and merging algorithm was reported in the 1950s and attributed to von Neumann \cite{kahn1951estimation}.
In 1996, Huber and Kim~\cite{huber1996weighted} 
modified von Neumann's original approach by grouping the trajectories into bins
and applying splitting and merging in each of the bins separately, thus preserving the bin weights.
It was recently shown that WE with fixed bin weights leads to convergent estimates\cite{aristoff2022ergodic,webber2020splitting}; in contrast, von Neumann's original method can be unstable in the large time limit\cite{aristoff2022ergodic,webber2020splitting}.

Attention to WE has grown, and the method is now implemented in the widely used WESTPA software package  \cite{zwier2015westpa,russo2021westpa} and the more recent wepy package \cite{lotz2020wepy}.  A Julia language package, used here, is also available, \cite{WEjl}.
Meanwhile, a growing community of researchers is promoting the method and developing it in different directions \cite{zhang2007efficient,dickson2014wexplore,adelman2011simulations,zwier2011efficient,abdul2014awe,ahn2021gaussian,ray2020weighted,ray2021markovian,ojha2022deepwest}.

In modern biochemical applications, WE is often used to estimate the MFPT for a stochastic dynamics
to transition from a source state $A$ into a target state $B$ \cite{zuckerman2017review}.
In addition, WE has been used to estimate other transition path statistics, including the distribution of reaction times from $A$ to $B$ \cite{zhang2007efficient,zwier2011efficient},
the distribution of entry points into $B$ \cite{bhatt2011reversibility,dickson2014wexplore},
and the characteristics of paths leading from $A$ to $B$ \cite{zhang2007efficient, adelman2011simulations}.
Here, we focus on the estimation of MFPTs,
which is a significant and challenging application because the inverse of the MFPT is the reaction rate constant \cite{bhatt2010steady}.

Other computational techniques for estimating MFPTs include Markov state models \cite{swope2004describing}, forward flux sampling \cite{allen2005sampling}, adaptive multilevel splitting \cite{cerou2007adaptive}, exact milestoning \cite{bello2015exact,aristoff2016mathematical}, non-equilibrium umbrella sampling \cite{warmflash2007umbrella}, and transition interface sampling \cite{van2003novel}.
Like WE, these approaches use large numbers of short, unbiased trajectories to compute the MFPT.
However, the following combination of features makes WE especially attractive:
\begin{itemize}

\item WE only requires simulation of the stochastic model forward in time, never backward.  
Thus, WE can be applied to a wide variety of stochastic models arising in chemistry \cite{huber1996weighted,donovan2013efficient} and in other areas, e.g., astronomy \cite{abbot2021rare}, climate science \cite{finkel2021learning,finkel2021exploring}, and systems biology \cite{donovan2013efficient,donovan2016unbiased}.

\item WE is fully parallelizable over time intervals of length $\tau$ between splitting/merging events \cite{darve2012computing,zwier2015westpa,russo2021westpa}.
Since the trajectories are simulated for the same time interval,
the parallellization can be managed simply and efficiently.

\item WE provides asymptotically unbiased, convergent estimates in the limit of many time steps for any choice of parameters \cite{zhang2010weighted, aristoff2022ergodic,webber2020splitting}.

\item WE has been observed to provide better MFPT estimates than estimates from Markov state models \cite{feng2015comparison} or non-equilibrium umbrella sampling \cite{adelman2013simulating}.
\end{itemize}

Despite progress over the last three decades,
theoretical questions 
about WE have remained 
unanswered until recently.
WE performance is 
highly dependent on the choice of parameters \cite{zhang2007efficient}, including the definition of the bins and the desired number of trajectories in each bin.
For a long time it was unclear what the optimal parameter choices would be.

Recent work in the mathematical literature sheds new light on the optimal parameter choices for WE \cite{aristoff2016analysis,aristoff2020optimizing,aristoff2022ergodic,webber2020splitting}.
The optimal merging coordinate is the local MFPT to $B$ given the current state. 
The optimal splitting coordinate is the local variance of the MFPT to $B$, given the current state and the time interval $\tau$.
There is a theoretical limit on the variance reduction achievable through WE,
and bins based on the 
optimal merging and splitting coordinates
ensure the optimal variance in the limit of many trajectories and many time steps.

In this article, we aim to communicate these recent mathematical advances in a brief, accessible way.
We identify the optimal splitting and merging coordinates in one- and two-dimensional examples.
We propose optimized bins based on these coordinates, which differ from the more traditional bins
based on the root mean square displacement (RMSD) to $B$. 
Our examples show that binning based on the RMSD can give catastrophically wrong results.
In such cases, more effective bins are needed, and optimization based on reaction coordinates is helpful.

The rest of the article is organized as follows.
Section \ref{sec:hill} discusses the computation of mean first passage times, which is a major
problem  addressed by WE;
Section \ref{sec:we} introduces the WE method;
Section \ref{sec:optimalCoords} identifies the optimal merging and splitting coordinates for WE; 
Section \ref{sec:optimalWE} recommends variance reduction strategies for WE;
and Section \ref{sec:conclusion} concludes.

\begin{table}[ht]
    \caption{Definitions of symbols used in this work.}
    \label{tab:symbol-definitions}
    \centering
    \begin{tabular}{c | @{\hspace{1em}}l@{}}
        Symbol & Definition \\ [0.5ex] \hline
        $X_t$ & underlying Markovian dynamics  \\
        $\beta$ &  inverse thermal energy ($1/k_B T$) \\
        $U$ & potential energy \\
        $\tau$ & evolution time interval or lag time \\
        $A$ & initial (source) set \\
        $B$ & target (sink) set \\
        $\rho_A$ & initial (source) distribution inside $A$ \\
        $T_B$ & first passage time to $B$ \\
        $N_t$ & number of arrivals in $B$ by time $t$ \\
        $q$ & committor function from $A$ to $B$ \\
        $J$ & steady-state flux into $B$ \\
        $\pi$ & steady state of source-sink dynamics \\
        $\langle \rangle$ & mean or average for source-sink dynamics \\
        $h$ & (flux) discrepancy function 
        \\ $v^2$ & (flux) variance function  \\
        $\hat{J}_t$ & WE estimate of steady-state flux up to time $t$ \\
        $\ppi$, $\hh$, $\vv$ & $\tau \to 0$ limits of $\pi$, $h$, $v^2$
    \end{tabular}
\end{table}

\section{Mean first passage times and the Hill relation} \label{sec:hill}

We will study the MFPT of a Markovian stochastic 
dynamics $X_t$ from a source state $A$ to a sink state $B$. 
In the biochemistry context,
$X_t$ could represent stochastic molecular dynamics such as Langevin dynamics or constant-energy dynamics generated with a stochastic thermostat. The MFPT could correspond to the characteristic time for folding, binding, or conformational change of a simulated protein.

To define the MFPT precisely, we must specify the distribution of starting points $\rho_A$ within the the source state
$A$.
Different choices of $\rho_A$ will lead to different MFPTs, even if $A$ remains the same. 
The distribution $\rho_A$ fully determines the MFPT,
which is defined as the averaged length of the trajectories initiated from the source distribution $\rho_A$
and absorbed upon reaching $B$.

When computing the MFPT, we ``recycle'' $X_t$ according to the source distribution $\rho_A$ upon arrival at the sink state $B$.
This means that the location immediately changes from $B$ to $A$, but the time index $t$ continues to increase as usual. 
We assume that the distribution of trajectories for this source-sink recycling dynamics converges as $t \rightarrow \infty$ to a unique steady-state distribution, denoted by $\pi$.

Under these recycling boundary condition,
the \emph{Hill relation} \cite{hill2005free} expresses the $\mfpt$ as the inverse of the 
steady-state flux $J$ of $X_t$ into $B$. To state this 
more precisely, 
write $T_B$ for the first passage time to $B$, and $N_t$ for the number of arrivals of $X_t$ in $B$ by time $t$.
The Hill relation then states:
\begin{align}
    \langle T_B\rangle_A &=\textup{ MFPT of }X_t \textup{ from }A \textup{ to }B \\
    \label{mfpt-long}
    &= 1/J = \left( \textup{steady-state flux into }B\right)^{-1} \\
    &=  \left(\frac{d}{dt}\langle N_t\rangle_{\pi}\right)^{-1}.
\end{align}
Here and elsewhere, we use subscripts to indicate the starting distribution for the dynamics $X_t$. 
Thus, $\langle T_B \rangle_A$ indicates the mean first passage time when $X_0$ is started from $\rho_A$, and $\langle N_t \rangle_{\pi}$ indicates the mean number of arrivals when $X_0$ is started from $\pi$.
Equation~\eqref{mfpt-long} 
holds for any $t\ge 0$, since $\pi$ is the steady 
state.

The application of the Hill relation transforms the computation of a long 
expected time (the MFPT) into 
a computation of a small 
rate (the steady 
state flux into $B$). 
This transformation makes it possible, in principle, to compute the MFPT using trajectories with lengths much shorter than the MFPT. 
Moreover, these short trajectories can be run in parallel to reduce the wall-clock time. 
Yet, a large number of trajectories are required since the flux can be very small.

Fortunately the Hill relation can be combined with enhanced sampling approaches like WE to help accurately estimate the small flux into $B$.
It is necessary that the trajectories approximate the steady-state distribution $\pi$,
but this can frequently be attained,
because the timescale for relaxation to steady state can be much shorter than the MFPT in problems of interest\cite{ben2007spectral}.

\begin{figure}
\includegraphics[width=\columnwidth]{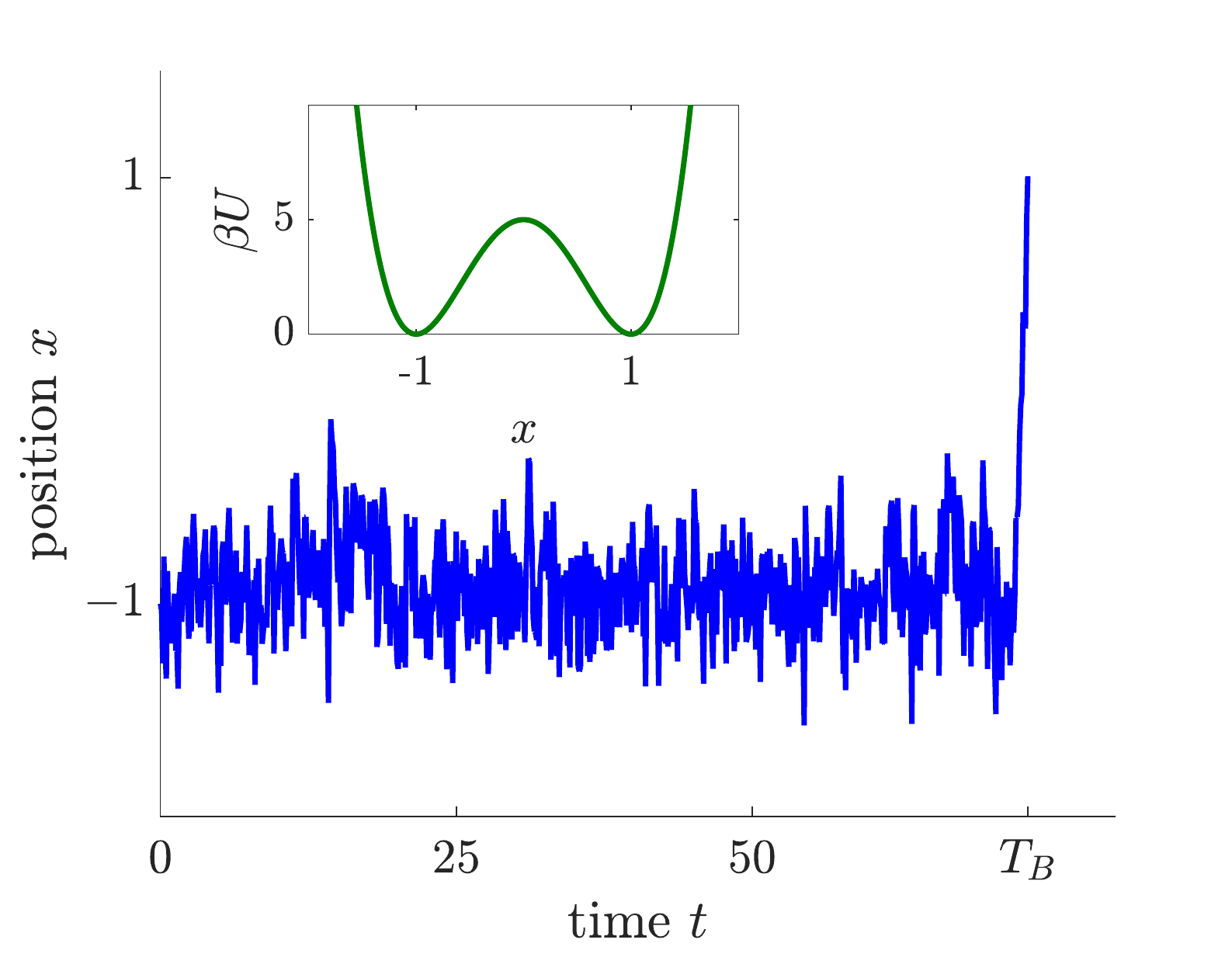}
\vskip-5pt
\caption{
\label{fig:rare_events}
{\em A first passage event out of a local minimum of $U$.}
After a long waiting period, the first entry into the set $B = \{x > 1\}$ occurs at time $T_B$.
The dynamics satisfies Smoluchowski dynamics \eqref{eq:overdamped} with 
$\beta U(x) = 5 (x-1)^2(x+1)^2$ and $D = 1/5$ (inverse time units).}
\end{figure}

Figure \ref{fig:rare_events} shows a simple system in which the steady-state flux from $A$ into $B$ is small.
Like many of the examples to follow, the
data in the figure comes from overdamped Langevin (``Brownian'')
dynamics
\begin{equation}
\label{eq:overdamped}
    \mathop{d X_t} = \Bigl[ \nabla D(X_t) - \beta D(X_t) \nabla U(X_t) \Bigr] \mathop{dt} + \sqrt{2 D(X_t)} \mathop{dW_t}
\end{equation}
associated with the Smoluchowski equation
\begin{equation}
\label{eq:Smoluchowski}
\partial_t p(x,t) = \nabla \cdot \Bigl[ D(x) \Bigl( \beta \nabla U(x) p(x,t) + \nabla p(x,t) \Bigr) \Bigr].
\end{equation}
Here, $\beta > 0$ is an inverse temperature constant,
$U: \mathbb{R}^n \rightarrow \mathbb{R}$ is a scalar potential, $D: \mathbb{R}^n \rightarrow \mathbb{R}$ is a diffusion coefficient,
${W}$ is a standard Brownian motion, and $p(x, t)$ is the probability density at location $x$ at time $t$.
For simplicity, we assume a dimensionless configurational coordinate $x$, which results in a diffusion coefficient $D$ exhibiting units of inverse time.

For exposition, many of our examples are in one or two spatial dimensions.  We emphasize, however, that the WE method and the core mathematical analysis here apply to any Markovian stochastic dynamics in any spatial dimension.

The continuous-time Smoluchowski equation needs to be discretized for numerical simulation. Therefore, we assume that $X_t$ is evolved as a discrete time series $X_0,X_\tau,X_{2\tau},\ldots$, and $X_t$ is only recycled when it occupies state $B$ at a multiple of the time interval $\tau$. The difference between discrete- and continuous-time flux into $B$ will be small if $\tau$ is small 
or the trajectories in $B$ are very slow to escape. 
We reserve variables $t$ and $\tau$ for the ``physical 
time'', i.e., the time index for the original dynamics $X_t$.

\section{Weighted ensemble: Splitting and merging} \label{sec:we}

\begin{figure}
\includegraphics[width=\columnwidth]{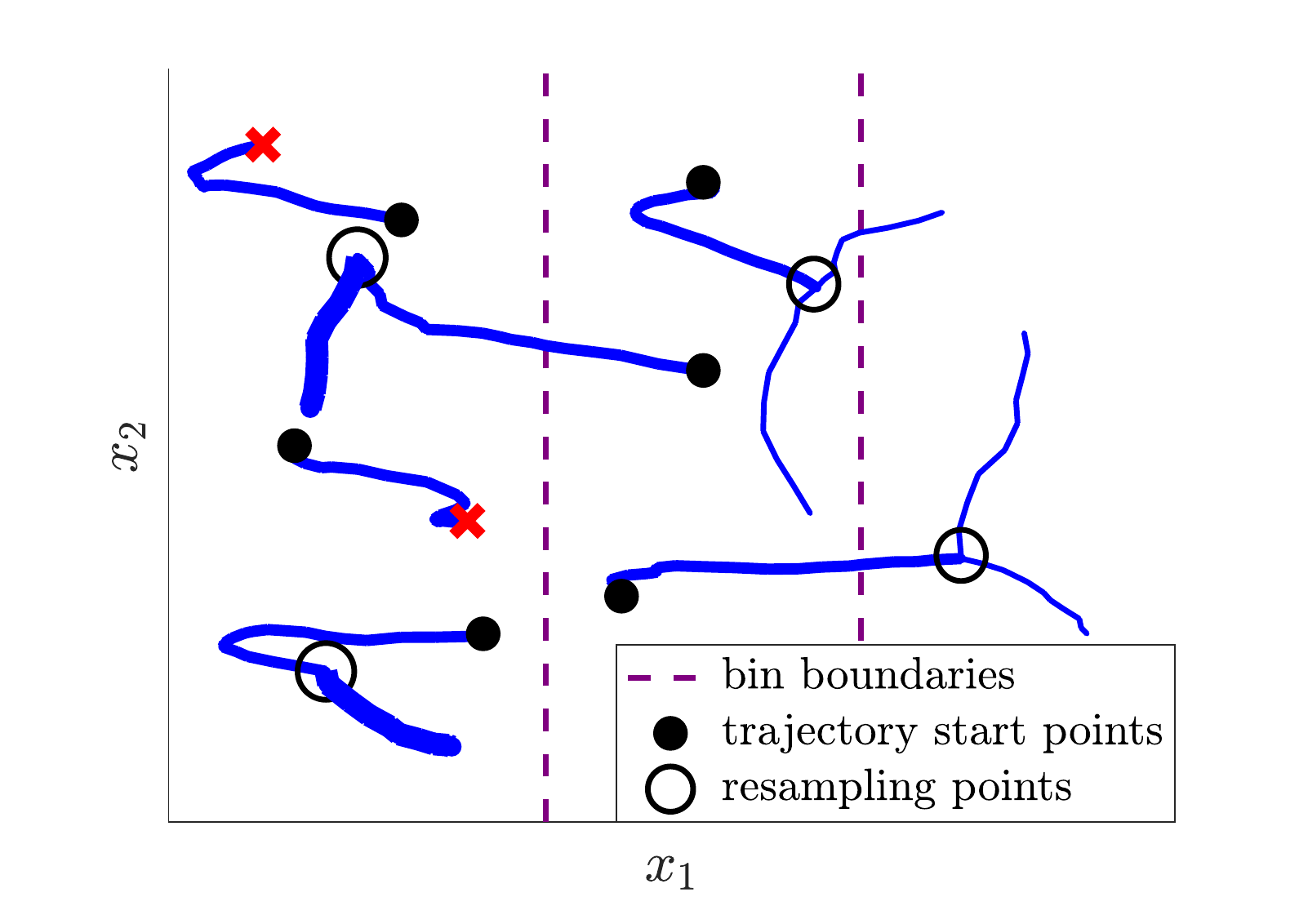}
\caption{\label{fig:WE_splitting_merging} 
{\em An example of WE evolution and resampling.}
Pictured 
are $6$ WE trajectories,
which undergo evolution, resampling, and another round of evolution.
The bins are based on the coordinate $x_1$.
During evolution steps, the trajectories travel between bins,
and the resampling enforces $2$ copies in each bin.
Circles indicate trajectories that are continued and possibly replicated during resampling;
crosses indicate trajectories that are pruned during resampling.
Trajectory weight is represented by line width.}
\end{figure}

Throughout this article we consider a simple version of WE in which the splitting and merging of trajectories are reformulated as {\em resampling}~\cite{zhang2010weighted,aristoff2020optimizing,aristoff2022ergodic,webber2020splitting}.
For a more general discussion of WE, see the review~\cite{zuckerman2017review}. 
The simple WE algorithm is described 
below and illustrated in Figure~\ref{fig:WE_splitting_merging}.

\begin{itemize}
    \item[1.] \textbf{Initialization}. Select starting points for $N$ trajectories. 
    Assign a weight to each trajectory so that the weights sum to one.
    
    \item[2.] \textbf{Resampling}.
    Partition the $N$ trajectories into {\em bins}, and select the desired number of copies for each bin, called the {\em allocation}. The bins are simply groupings of the trajectories, and they may change in time. The allocations are positive integers which may change with time but are assumed to always sum to $N$.
    
    Next, \emph{resample}
    trajectories within each bin with probabilities proportional to the weights, and divide the total weight in each bin evenly among the resampled trajectories~\cite{aristoff2022ergodic}.

    \item[3.] \textbf{Dynamics}. Evolve the trajectories independently for time $\tau$
    according to the underlying dynamics.

    \item [4.] \textbf{Convergence}. Repeat steps 2-3 as long as desired or possible. 
    Estimate the steady-state flux $J$ using the average WE probability flux entering $B$ from the burn-in time $t_0$
    to the final time $t$. That is,
    \begin{equation}
        \hat{J}_t =\frac{\textup{Total WE weight entering $B$}}{t - t_0},
        \label{flux_we}
    \end{equation}
    where $t-t_0$ is the physical time of simulations following the burn-in.
    The estimate for $J$ in turn yields a MFPT estimate via the Hill relation \eqref{mfpt-long}.
\end{itemize}

The crucial step in WE is the resampling step, which leads to the splitting and merging of trajectories. Splitting is the dynamics of replicating trajectories  that are ``valuable,'' while merging is the act of combining trajectories that  are ``similar.'' Splitting improves the accuracy of the flux estimates, especially when the duplicated trajectories are expected to make large contributions to the future flux. With repeated splitting, merging is needed to limit the number of trajectories.

Although splitting and merging are traditionally applied separately, in the simple WE algorithm above these steps are combined via {\em resampling}\cite{zhang2010weighted}.
In each bin, trajectories are resampled with probabilities proportional to their weights, with the resampled trajectories each receiving an equal proportion of the bin weight.  
Resampling is easier to formulate and implement than the traditional splitting and merging.
Moreover, resampling  is \emph{optimal} when trajectories within a given bin are, except for their weights, indistinguishable~\cite{darve2012computing}.

Splitting and merging must be balanced, and a main contribution of our recent work~\cite{aristoff2016analysis,aristoff2020optimizing,aristoff2022ergodic,webber2020splitting} has been identifying the  regions of state space 
in which splitting or merging should be promoted.
Identifying such regions naturally leads to variance reduction strategies.
We discuss these developments in Sections \ref{sec:optimalCoords} and \ref{sec:optimalWE}.  

Given sufficient resources, the WE method terminates when the estimated flux
converges to a near-constant value.
However, if steady state cannot be reached during the available simulation time, enhanced methods for WE initialization are needed.
For example,
incorporating an appropriate burn-in time $t_0 > 0$ can reduce transient relaxation effects \cite{adhikari2019computational},
and adjusting the initial 
weights can also improve  convergence~\cite{copperman2020accelerated}.

WE can be used to estimate other transition path statistics in addition to the steady-state flux. For instance, the WE-generated paths and the associated weights can be used to estimate the distribution of reaction times (i.e., event durations) and the distribution of mechanistic pathways \cite{zhang2007efficient}.
See Appendix~\ref{app:general} for a discussion.  

\section{Optimal splitting and merging coordinates}
\label{sec:optimalCoords}

Recent mathematical analysis has revealed the existence of optimal reaction coordinates for merging and splitting in WE
\cite{aristoff2020optimizing,webber2020splitting,aristoff2022ergodic}.
Sections \ref{sec:optimal_fns} and \ref{sec:maximum_gain} provide general formulas for these optimal reaction coordinates that assume no particular 
type of dynamics or dimensionality,
while Section \ref{sec:smoluchowski} provides more explicit formulas that hold for the Smoluchowski dynamics.

\subsection{Optimal merging and splitting coordinates}
\label{sec:optimal_fns}

\textbf{Optimal merging}.
Merging trajectories, i.e., pruning some members of a chosen set, is least harmful and most beneficial 
when the groups of 
trajectories to be merged are in some 
sense ``similar.'' For the MFPT problem, the scalar reaction coordinate 
that characterizes this similarity is
the {\em flux discrepancy function} \cite{aristoff2020optimizing}
\begin{equation}
    h(x) = \lim_{t \to \infty} \left[\left\langle N_t \right\rangle_x - \left\langle N_t \right\rangle_{\pi} \right].
    \label{discrep}
\end{equation}
$N_t$ counts the total number of crossings into $B$ at times $s =\tau, 2\tau, \ldots, t$.
Thus, $h(x)$ is 
the difference in expected future flux between trajectories started at the particular point $x$ and trajectories started from the steady-state distribution $\pi$.
Since trajectories with similar $h$ values make similar expected contributions to the flux estimate, merging 
trajectories with similar $h$ values is appropriate~\cite{aristoff2020optimizing,aristoff2022ergodic}. 

Using the Hill relation, the flux discrepancy function can be rewritten in terms of MFPTs initiated from different starting distributions.
As shown in Appendix \ref{sec:loc-mfpt},
\begin{equation}\label{eq:h}
    h(x) =  \frac{\langle T_B \rangle_{\pi} - \langle T_B\rangle_x}{ \langle T_B\rangle_A},
\end{equation}
where we recall $T_B$ is the first passage time into $B$ excluding time $t = 0$.
Equation
\eqref{eq:h} allows us to reinterpret the optimal merging function as a normalized difference between the local MFPT to $B$ starting from $x$ and the MFPT for trajectories initiated from the steady state $\pi$.
Identity \eqref{eq:h} holds independently of both the dimension and the type of Markovian dynamics.
It shows that the flux discrepancy function and the local MFPT 
are equivalent optimal coordinates 
for merging. 

\textbf{Optimal splitting}.
Splitting -- the duplication of certain WE trajectories  -- increases local sampling and is of high benefit in regions that 
are important for the flux computation.
We now introduce the scalar reaction coordinate that describes optimal splitting behavior,
the {\em flux variance function} \cite{aristoff2020optimizing}
\begin{equation}
    \label{eq:vdt}
    v(x)^2 = \tau^{-1} \textup{Var}_x \bigl[ h_0(X_\tau) \bigr],
\end{equation}
where $h_0$ is a version of the discrepancy function that counts flux at time $t = 0$,
i.e., $h_0(x) = \mathds{1}_B(x) + h(x)$, and $\textup{Var}_x$ denotes variance for the source-sink dynamics started at $x$.
The flux variance function quantifies the variation
in the expected flux over a time interval $\tau$ for trajectories initiated at $x$.

As a heuristic strategy, a favorable allocation should satisfy~\cite{aristoff2016analysis,aristoff2020optimizing,aristoff2022ergodic}
\begin{equation}\label{eq:opt_alloc_rule}
    \#\textup{ trajectories near }x \propto \pi(x)v(x).
\end{equation}
Allocating in this way minimizes the future 
contribution to the flux variance, assuming that
WE is in steady state~\cite{aristoff2016analysis,aristoff2020optimizing,aristoff2022ergodic}.
We refer to this as the {\em optimal allocation distribution} and provide an intuitive derivation in Appendix \ref{app:gen_var}.

As a more rigorous mathematical result, the following establishes the optimality of $h$ and $v$ as WE reaction coordinates.

\emph{Theorem}~\cite{webber2020splitting}. 
If the recycled dynamics $X_t$ is geometrically ergodic, then the WE method satisfies the following:
\begin{enumerate}
\item For any choice of bins and allocations,
the WE flux estimate $\hat{J}_t$ given in \eqref{flux_we} converges with probability one to the inverse MFPT,
\begin{equation}
    \lim_{t \rightarrow \infty} \hat{J}_t = \frac{1}{\langle T_B\rangle_A}.
\end{equation}
\item 
For any $\epsilon > 0$ and any choice of bins and allocations, the WE variance satisfies
\begin{equation}
    \textup{Var}\, \hat{J}_t
    \geq \frac{1 - \epsilon}{Nt} \Bigl(\int v(x) \pi(x) \mathop{dx}\Bigr)^2
\end{equation}
when $t$ is sufficiently large.
\item For any $\epsilon > 0$, 
if the bins are sufficiently small rectangles in $h$ and $v$ coordinates
and if the allocation for each bin is proportional to
$\int_{\textup{bin}} v(x) \pi(x) \mathop{dx}$,
the WE variance satisfies
\begin{equation}
\label{eq:near_optimal}
    \textup{Var}\, \hat{J}_t
    \leq \frac{1 + \epsilon}{Nt} \Bigl(\int v(x) \pi(x) \mathop{dx}\Bigr)^2
\end{equation}
for sufficiently large $t$ and $N$.
\qedsymbol
\end{enumerate}

The theorem holds regardless of dimension, temperature, time $\tau$, and type of Markovian  dynamics. It highlights the following WE strategy that is optimal in the limit of large $N$ and $t$: trajectories with similar $h$ and $v$ values are merged, while splitting enforces $\textup{Const}\times\pi(x)v(x)$ trajectories near $x$.
The theorem supports the interpretation~\eqref{eq:opt_alloc_rule} of $\pi(x)v(x)$ 
as the optimal distribution for trajectory 
allocation.

\subsection{Maximum gain of WE over direct dynamics}
\label{sec:maximum_gain}

The above mathematical theory enables a quantitative comparison between WE and direct Monte Carlo sampling of first-passage events.
Here, we define direct Monte Carlo as independent trajectories without any splitting or merging.

Assuming that trajectories are started from the steady state $\pi$,
the direct Monte Carlo variance with $N$ trajectories can be written~\cite{webber2020splitting}
\begin{equation}\label{eq:direct_MC}
    \frac{1}{Nt} \int v(x)^2 \pi(x) \mathop{dx}.
\end{equation}
This formula reveals that the flux variance function $v^2$ is intrinsic to direct Monte Carlo dynamics as well as WE dynamics, since $v^2$ quantifies the variance in the flux estimates as trajectories are evolved forward over a time interval $\tau$.
Dividing \eqref{eq:direct_MC} by~\eqref{eq:near_optimal}, the ratio of 
direct Monte Carlo variance 
to optimal WE variance is then~\cite{webber2020splitting}
\begin{equation}
\begin{split} \label{eq:opt_var_const}
    &\textup{Optimal gain over direct Monte Carlo}\\
    &\quad= \frac{\int v^2(x)\pi(x)\,dx}{\left(\int v(x)\pi(x)\,dx\right)^2}.
\end{split}
\end{equation}
This ratio quantifies the maximal possible variance reduction achievable by WE in the large time limit, for any number of trajectories.

The optimal gain over direct Monte Carlo has important implications.
Analytical and numerical investigation of this quantity can yield insight into how much benefit is possible from using WE. 
Additionally, knowing the optimal gain over direct Monte Carlo as a reference enables quantitative comparisons of different WE binning strategies against the theoretically optimal performance.

\subsection{Optimal WE for one-dimensional Smoluchowski dynamics}
\label{sec:smoluchowski}

To make the preceding theory more explicit,
we consider Smoluchowski (overdamped Langevin) dynamics in a one-dimensional setting  with a source state $A = \{x \leq a\}$ and a sink state $B = \{x \geq b\}$, where $a < b$.
The recycling distribution is a delta function at the edge $a$ of $A$, i.e., $\rho_{A} = \delta(x - a)$. We consider the continuous-time limit, $\tau \rightarrow 0$, which leads to simpler, more interpretable mathematical expressions.

\begin{figure}
\centering
    
\includegraphics[width=\columnwidth]{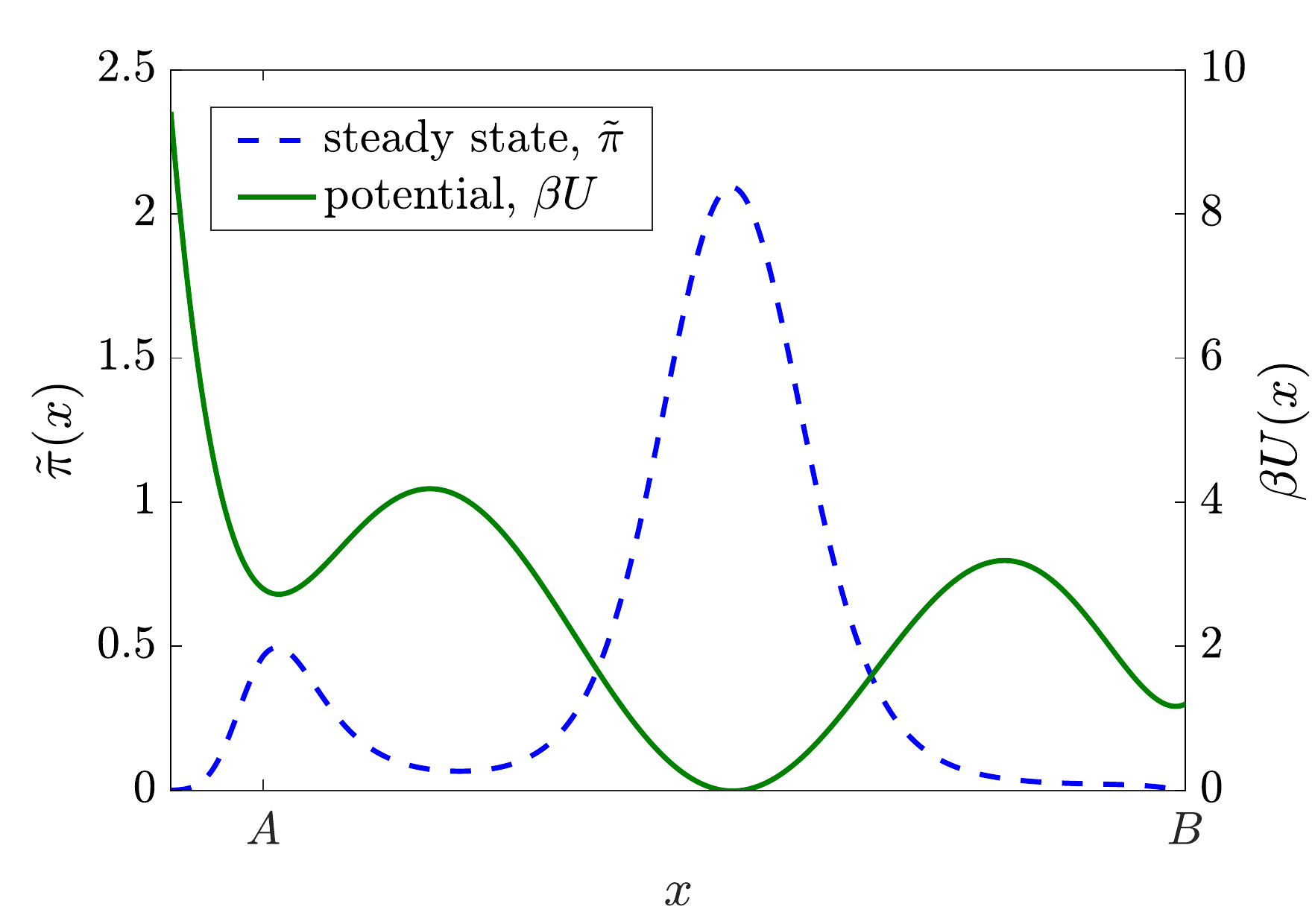}
     \caption{{\em The steady state of the source-sink dynamics.} Plotted is the steady state $\ppi$ of the Smoluchowski dynamics with recycling from $B$ to $A$, a diffusion coefficient $D = 1$, and $\beta = 4$.
     In this and subsequent figures, $U(x) = 5(x-1)^2 x^2(x+1)^2 + 0.5 x^2 - 0.2x$.}
 \label{fig:pi}
\end{figure}

In the limit $\tau \rightarrow 0$,
the dynamics $X_t$ is observed at all times and recycling occurs immediately upon entry into $B$.
The steady-state distribution $\ppi$ can be calculated using the relation~\cite{lu2015reactive}
\begin{equation}
\label{eq:steady1}
    \ppi(x) \propto 
    [1-\qq(x)] e^{-\beta U(x)}.
\end{equation}
Here, $\qq$ is the {\em committor} function, the probability for $X_t$ to reach $B$ before $A$ starting from $x$, which is given by~\cite{gardiner1985handbook}
\begin{equation}
\label{eq:steady4}
\qq(x) = \begin{cases}
    0 &  x \leq  a, \\
    \dfrac{\int_a^x e^{\beta U(y)}/D(y)\,dy}
    {\int_a^b e^{\beta U(y)}/D(y)\,dy} & x<a<b,\\
    1 & x\geq b.
\end{cases}
\end{equation}
Using \eqref{eq:steady1} and \eqref{eq:steady4}, the steady-state distribution is supported on $\{x < b\}$, and it satisfies
\begin{equation}
\label{eq:pi}
 \ppi(x) 
 \propto e^{-\beta U(x)}\int_{\max\{x,a\}}^b \frac{e^{\beta U(y)}}{D(y)}\,dy.
\end{equation}
Compared to the Boltzmann density $\propto e^{-\beta U(x)}$,
the steady-state density $\tilde{\pi}$ removes mass from the regions near $B$ due to the recycling of trajectories.
See Figure \ref{fig:pi} for an illustration.

The flux discrepancy function $h = h_\tau$ and the flux variance function $v = v_\tau$ depend implicitly on the evolution interval $\tau$, and they converge to well-defined limits as $\tau \rightarrow 0$:
\begin{equation}
    \hh(x) = 
    \lim_{\tau \to 0} h_{\tau}(x),
    \qquad
    \vv^2(x) = \lim_{\tau \to 0} v_{\tau}(x)^2.
\end{equation}
As a result of \eqref{eq:h}, the discrepancy function $\hh$ satisfies
\begin{equation}
\label{eq:combo1}
    \tilde{h}(x) =  \frac{\langle \tilde{T}_B \rangle_{\tilde{\pi}} - \langle \tilde{T}_B\rangle_x}{ \langle \tilde{T}_B\rangle_A}.
\end{equation}
Here, $\tilde{T}_B$ is the exact MFPT function to the target state $B$, which is given for $x \leq b$ by~\cite{gardiner1985handbook}
\begin{equation}
\label{eq:tauB}
 \langle \TT_B\rangle_x = \int_x^b \frac{e^{\beta U(z)}}{D(z)}\int_{-\infty}^z e^{-\beta U(y)}\,dy\,dz.
\end{equation}
The variance function $\vv$ satisfies the relations
\begin{equation}
\label{eq:v}
    \vv^2(x) = \lim_{\tau \rightarrow 0} \frac{
    \textup{Var}_x \bigl[ \tilde{h}(X_{\tau}) \bigr]}{\tau} = 2D(x) \Bigl|\frac{d}{dx} \tilde{h}(x)\Bigr|^2,
\end{equation}
where the first equality follows from the definition \eqref{eq:vdt} and the second equality comes from applying Itô's lemma \cite{gardiner1985handbook}.
Formula \eqref{eq:v} suggests that there should be 
more splitting in regions of high $\tilde{h}$ variability, or, equivalently, high variability in the local MFPT $\langle \tilde{T}_B\rangle_x$.

\begin{figure}
\includegraphics[width=\columnwidth]{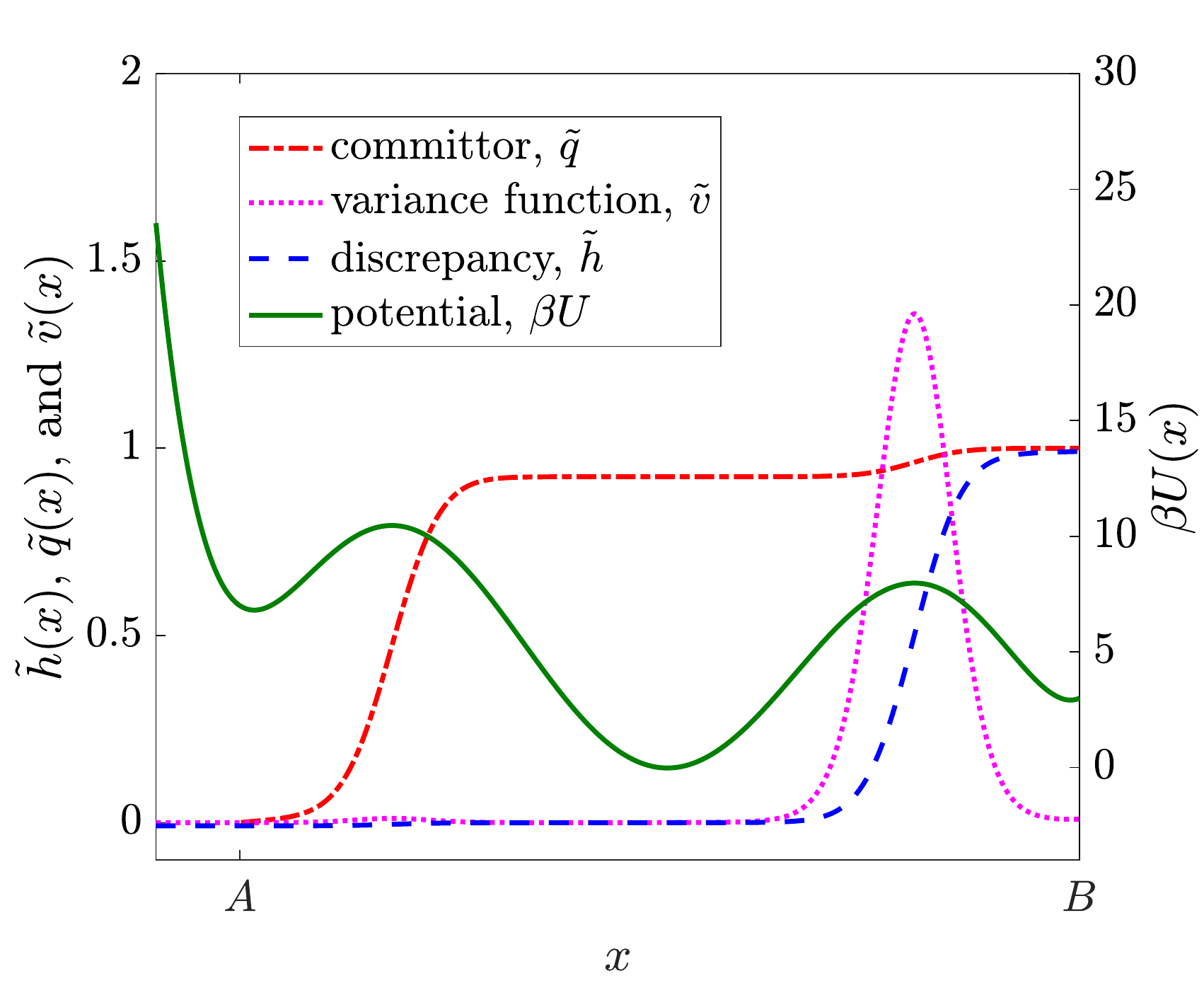}\vskip-5pt
\caption{\label{fig:h_vs_q} {\em The ideal  reaction coordinates for WE.}  Plotted are the discrepancy function $\hh$ the committor $\qq$ and the variance function $\vv$ for the Smoluchowski dynamics 
with recycling from $B$ to $A$, a diffusion coefficient $D = 1$, and $\beta = 10$.
}\label{fig:coordinates}
\end{figure}

Note that the committor $\tilde{q}$ is usually understood to be the relevant coordinate for computing
the MFPT and similar problems~\cite{makarov2020value}. We have shown, however, that two different scalar coordinates, namely $\tilde{h}$ and $\tilde{v}^2$, 
are the relevant ones for splitting 
and merging in the context of WE simulation. The difference between these coordinates is exhibited by a model problem with an asymmetric two-barrier system in Figure~\ref{fig:h_vs_q}.
For this problem, the committor $\qq$ would not be an ideal reaction coordinate since it poorly resolves the largest barrier on the forward path from $A$ to $B$.
In contrast, $\hh$ and $\vv$ resolve the largest forward barrier, making them appropriate for WE.

\begin{figure}
\centering
    
\includegraphics[width=\columnwidth]{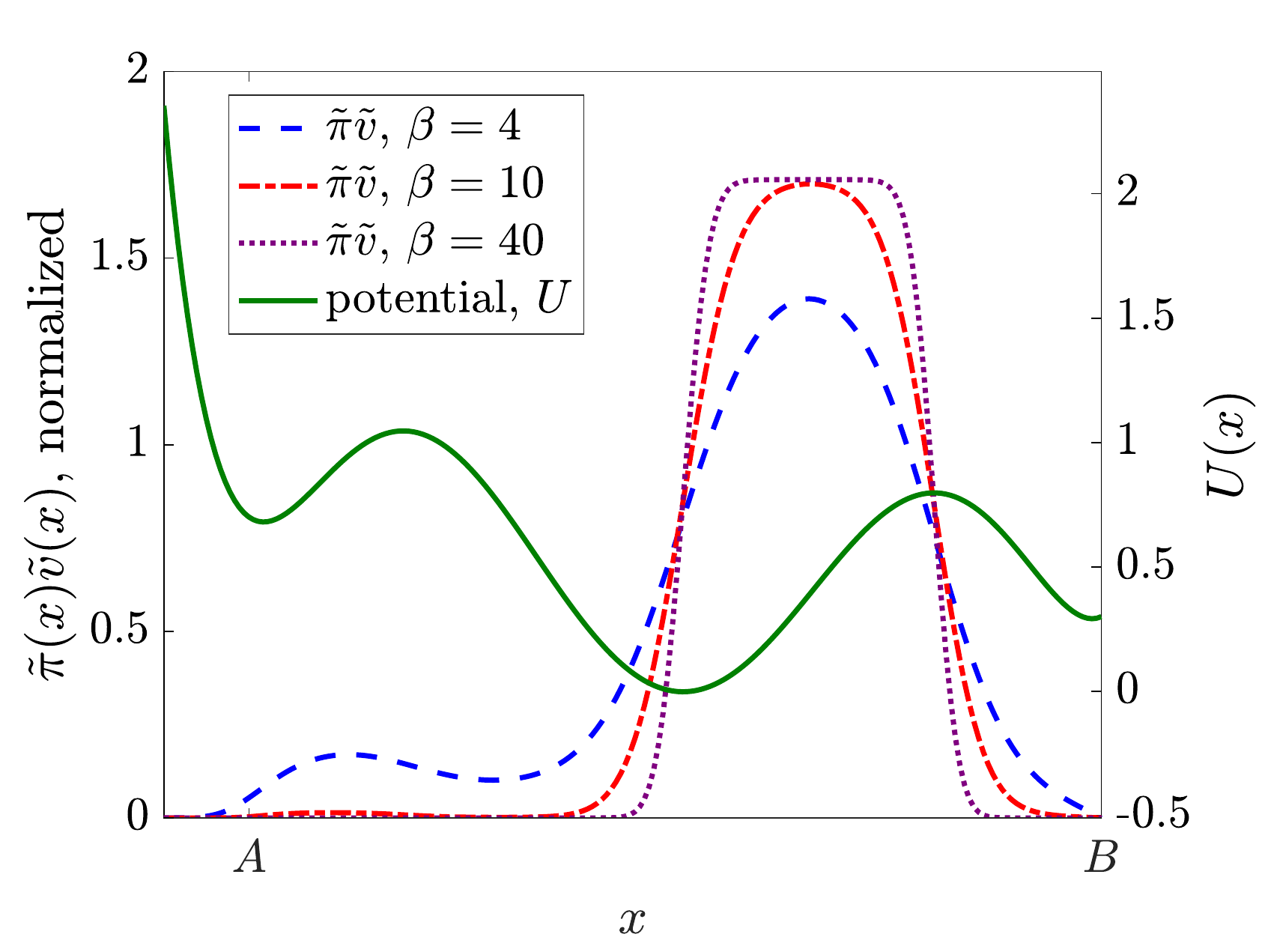}
     \caption{{\em The optimal allocation distribution at low temperature.} Plotted is the optimal allocation distribution $\ppi \vv$ for the Smoluchowski dynamics with recycling from $B$ to $A$ and with increasing values of $\beta$ when $D = 1$. 
     In the limit $\beta \rightarrow \infty$, the optimal allocation distribution is constant
     along the interval of largest energy increase; see~\eqref{eq:low_temp}.
     \label{fig:low_temp}}
    \end{figure}

Lastly, a simple formula for the optimal WE allocation holds in the low-temperature limit as $\beta \rightarrow \infty$.
Assume that the largest energy increase on the forward path to $B$ 
occurs over the interval $[x_-, x_+]$, i.e., 
\begin{equation}
    \Delta U = U(x_+) - U(x_-) = \max_{a \leq y \leq b,\, x < y} (U(y) - U(x)).
\end{equation}
Note that the interval $[x_-, x_+]$ can include multiple energy barriers.
Then, as $\beta \rightarrow \infty$, an application of Laplace's method (see Appendix \ref{app:low_temp}) yields
\begin{equation}
\label{eq:low_temp}
\frac{\ppi(x) \vv(x)}{\int \ppi(y) \vv(y) \mathop{dy}} \to 
\begin{cases}
    0 & x  \leq x_-,\\
    \dfrac{D^{-1 \slash 2}(x)}{\int_{x_-}^{x_+}
    D^{-1 \slash 2}(y) \mathop{dy}} & x_- < x < x_+,\\
    0 & x \geq x_+.
\end{cases}
\end{equation}
Thus, the 
optimal allocation is proportional to $1/\sqrt{D}$ over the interval $[x_-, x_+]$,
and it is vanishingly small elsewhere as illustrated in Figure \ref{fig:low_temp}.

 In the low-temperature limit $\beta \rightarrow \infty$, the optimal gain over direct Monte Carlo sampling grows exponentially with the size of the largest energy increase (see Appendix \ref{app:low_temp}): 
\begin{multline}
\label{eq:exponential}
    \textup{Gain over direct Monte Carlo} \\
    \overset{\beta\to \infty}{\sim} \frac{\pi/\beta}{\left(\int_{x_-}^{x_+} \sqrt{D(x_+) \slash D(x)}\,dx\right)^2}\frac{e^{\beta \Delta U }}{\sqrt{|U''(x_-)U''(x_+)|}}.
\end{multline}
This provides a formal explanation of earlier numerical findings~\cite{zhang2007efficient,degrave2018large} that WE's advantage over direct Monte Carlo grows dramatically as $\beta \Delta U$ increases.

\section{A principled WE strategy}\label{sec:optimalWE}

The theoretical results above can serve as guidelines for optimizing WE. 
For example, we describe one principled WE strategy called ``MFPT binning'' in Section \ref{sec:bin_opt},
and we compare MFPT binning to a more traditional WE binning strategy in Section \ref{sec:comp}.

\subsection{MFPT binning}
\label{sec:bin_opt}

\begin{figure}
\includegraphics[width=\columnwidth]{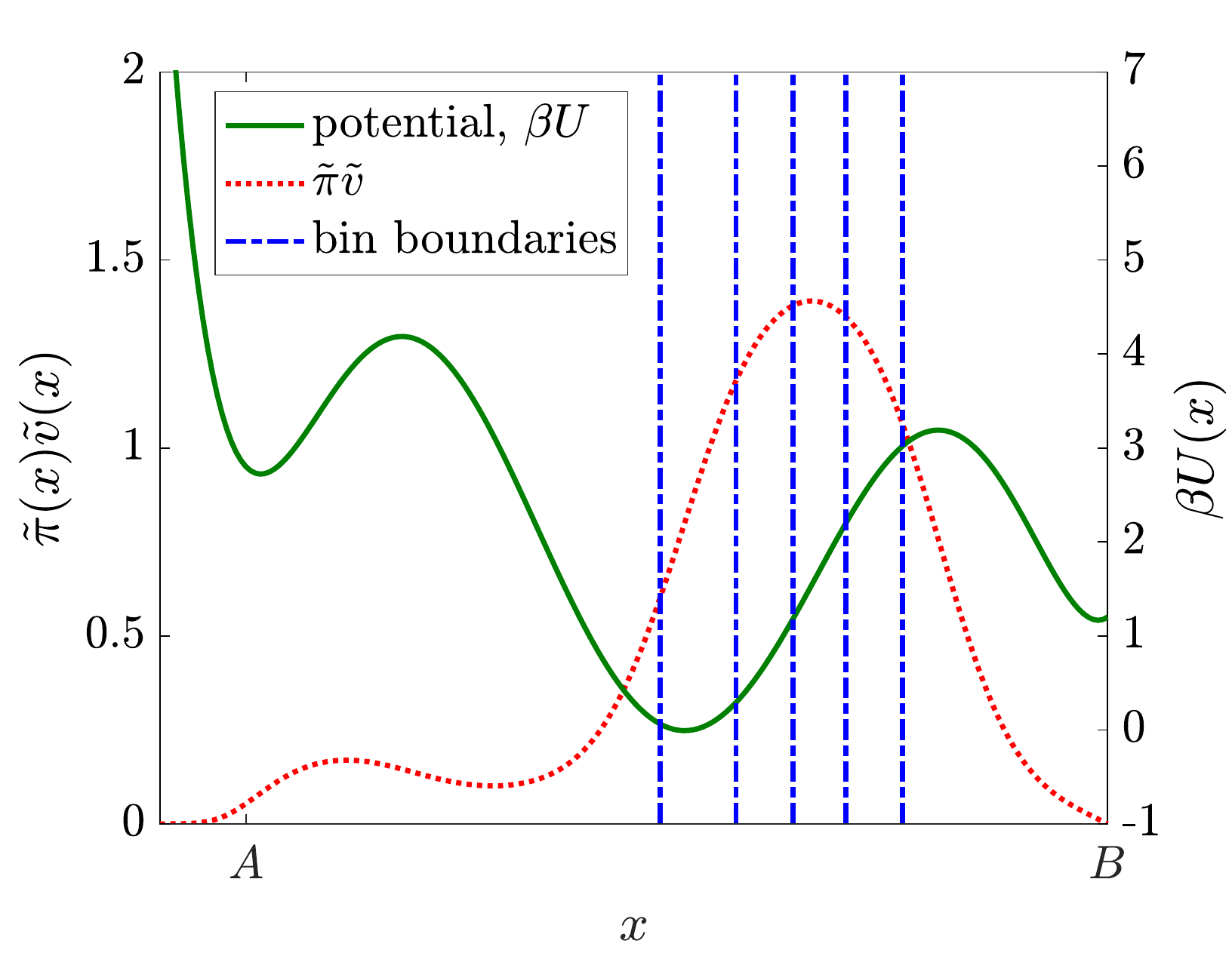}
\vskip-5pt
\caption{\label{fig:binning} 
{\em Illustration of MFPT binning strategy.} 
Pictured are $6$ bins, 
with dividing surfaces between bins indicated by the vertical lines.
Bin boundaries are chosen to make the integral of $\ppi \vv$ within each bin constant. Here, 
$D = 1$ and $\beta = 4$.
In higher-dimensional spaces, bin boundaries would be
level sets of $h$ as in Figure~\ref{fig:msm2d}. }
\end{figure}

From the theory in Section \ref{sec:optimal_fns},  merging has a low cost among trajectories with similar $h$ values. 
This motivates the following \emph{MFPT binning} strategy in which bins are comprised of similar $h$ values~\cite{aristoff2020optimizing,aristoff2022ergodic}, or equivalently based on \eqref{eq:h}, similar values of the local MFPT to $B$:

\begin{itemize}
    \item[1.] 
    Define the bins to be intervals in the $h$ coordinate.
    If there are $K$ bins,
    choose the endpoints $h_0< h_1<\ldots< h_K$ such that
    \begin{equation}
        \label{e:hint}
        \int_{h_k\leq h(x)\leq h_{k+1}} \pi(x)v(x) dx = \text{Const.}
    \end{equation}
    \item[2.] Allocate trajectories uniformly among these $K$ bins, so that approximately $N/K$ trajectories are assigned to each bin. 
\end{itemize}
This approach agrees with strategy~\eqref{eq:opt_alloc_rule} while also satisfying the traditional WE rule of keeping the number of trajectories per bin nearly constant.
See Figure~\ref{fig:binning} for an illustration. 

The MFPT binning strategy is similar to the one described in Section \ref{sec:optimal_fns}, which attains the lowest possible variance in the limit of many trajectories and time steps. However, instead of bins that are rectangles in $h$ and $v$ coordinates, we define bins as intervals of $h$ values alone.
This alteration still performs well in our model problem in Section \ref{sec:comp}, even with a small number of trajectories ($100$--$1000$) and just a few bins ($5$--$10$). 

The MFPT binning strategy requires an approximation to the flux discrepancy function $h$ and flux variance function $v$. 
Strategies for estimating the functions will be explored in future work.
We note that even crude estimates of $h$ and $v$ may be helpful, compared to naive or ad-hoc binning strategies, and WE remains unbiased regardless of the approximation quality of $h$ and $v$.
As a concrete strategy~\cite{aristoff2020optimizing}, initial simulations can be performed, and discrete Markov state models~\cite{chodera2014markov} (MSMs) can be used to approximate $h$ and $v$ based on the relationships to the MFPT described in Section~\ref{sec:optimalCoords}.

In contrast to this MFPT binning strategy, the most common WE approach~\cite{zuckerman2017review} involves bins defined using root mean square displacement (RMSD) to $B$ in some feature space, with the same number of trajectories allocated to each bin. This approach would be  optimal if the dynamics in the RMSD coordinate were well-described by the one-dimensional Smoluchowsi equation~\eqref{eq:Smoluchowski} with large $\beta$ and constant $D$; see Section \ref{sec:smoluchowski} and Figure~\ref{fig:coordinates}. However, the RMSD coordinate rarely behaves in this way for complex systems, and our results below show that RMSD binning can catastrophically fail when the RMSD coordinate is very different from the MFPT.

\subsection{Numerical comparison}
\label{sec:comp}

In this section, we numerically compare two strategies for binning and allocation within WE:
\begin{enumerate}
    \item The MFPT binning strategy discussed in Section \ref{sec:bin_opt}.
    \item Traditional WE bins based on RMSD to $B$ with allocations proportional to $\int_{\textup{bin}} v(x) \pi(x) \mathop{dx}$.
\end{enumerate}
In both strategies, the allocations enforce the heuristic \eqref{eq:opt_alloc_rule} of having $\appropto \pi(x) v(x)$ trajectories near $x$, which is derived mathematically in Section \ref{sec:optimalCoords}.

\begin{figure}
    \centering
    \includegraphics[width=\columnwidth]{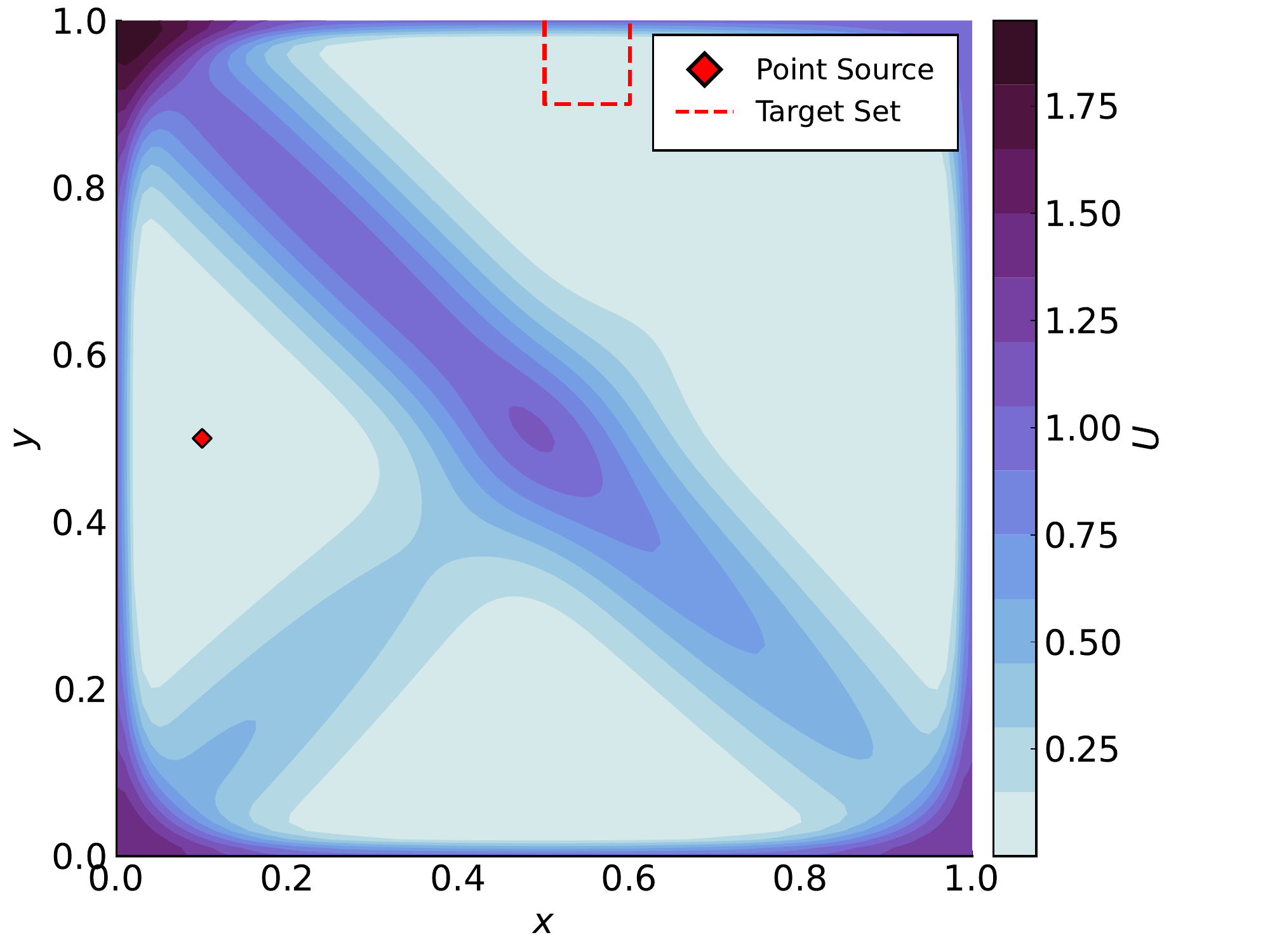}
    \includegraphics[width=.95\columnwidth]{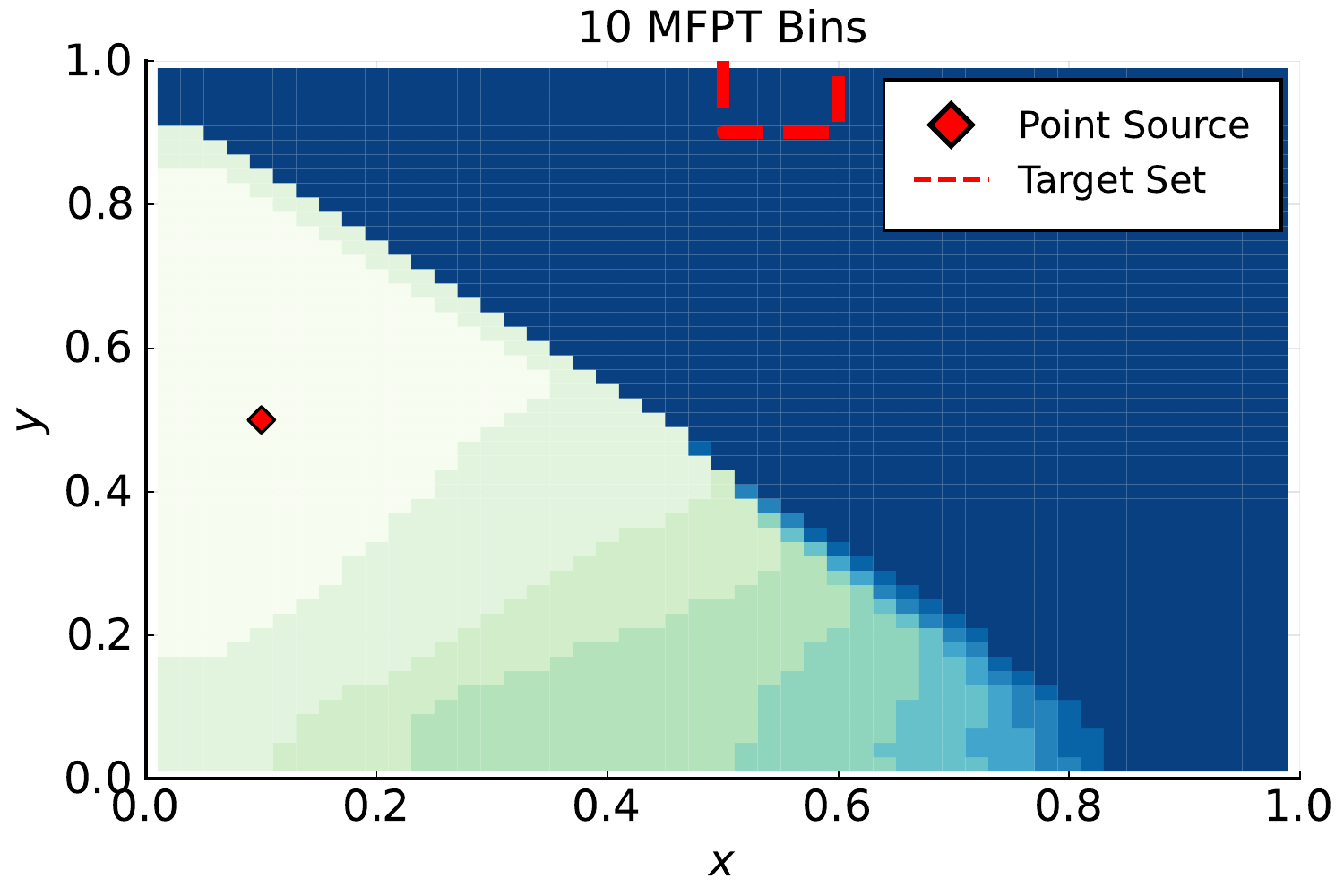}
    \includegraphics[width=.95\columnwidth]{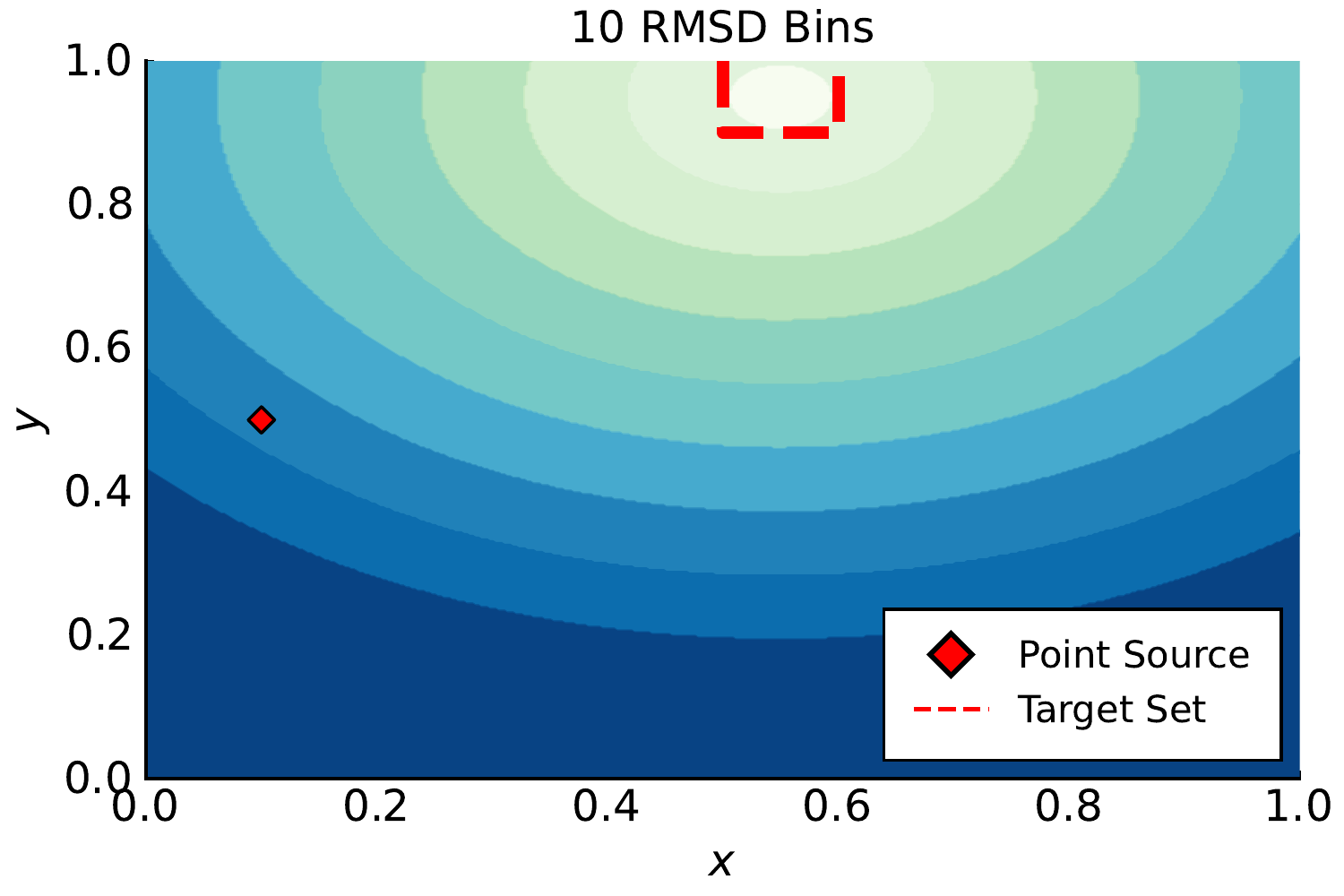}
    \caption{\emph{Different binning strategies in a challenging two-dimensional model.} 
    Pictured are the potential $U$ (top), MFPT bins (middle), and RMSD bins (bottom), with the source $A$ depicted as a filled dot and the sink $B$ depicted as a dashed rectangle.
    Different shades indicate different bins.
    The model is fully specified in Appendix \ref{app:sim_details}: see the potential \eqref{utwod}.}
    \label{fig:msm2d}
\end{figure}

We will apply these WE strategies to the two-dimensional Smoluchowski model system 
pictured in Figure \ref{fig:msm2d},
with full specifications in Appendix \ref{app:sim_details}.
We begin our study of this model by constructing a fine-grained MSM approximation of functions $\pi$, $h$, and $v$.
Next, we use these approximations to construct the MFPT bins in Figure \ref{fig:msm2d}.
Observe that the MFPT bins resolve the largest energy barriers on the forward path from $A$ to $B$, whereas the RMSD bins have concentric circle bin boundaries that do not account for the energy landscape at all.
See Appendix \ref{app:sim_details} for further bin construction details.  These computations were performed using \cite{WEjl}.

\begin{figure}
    \centering
    \includegraphics[width=\columnwidth]{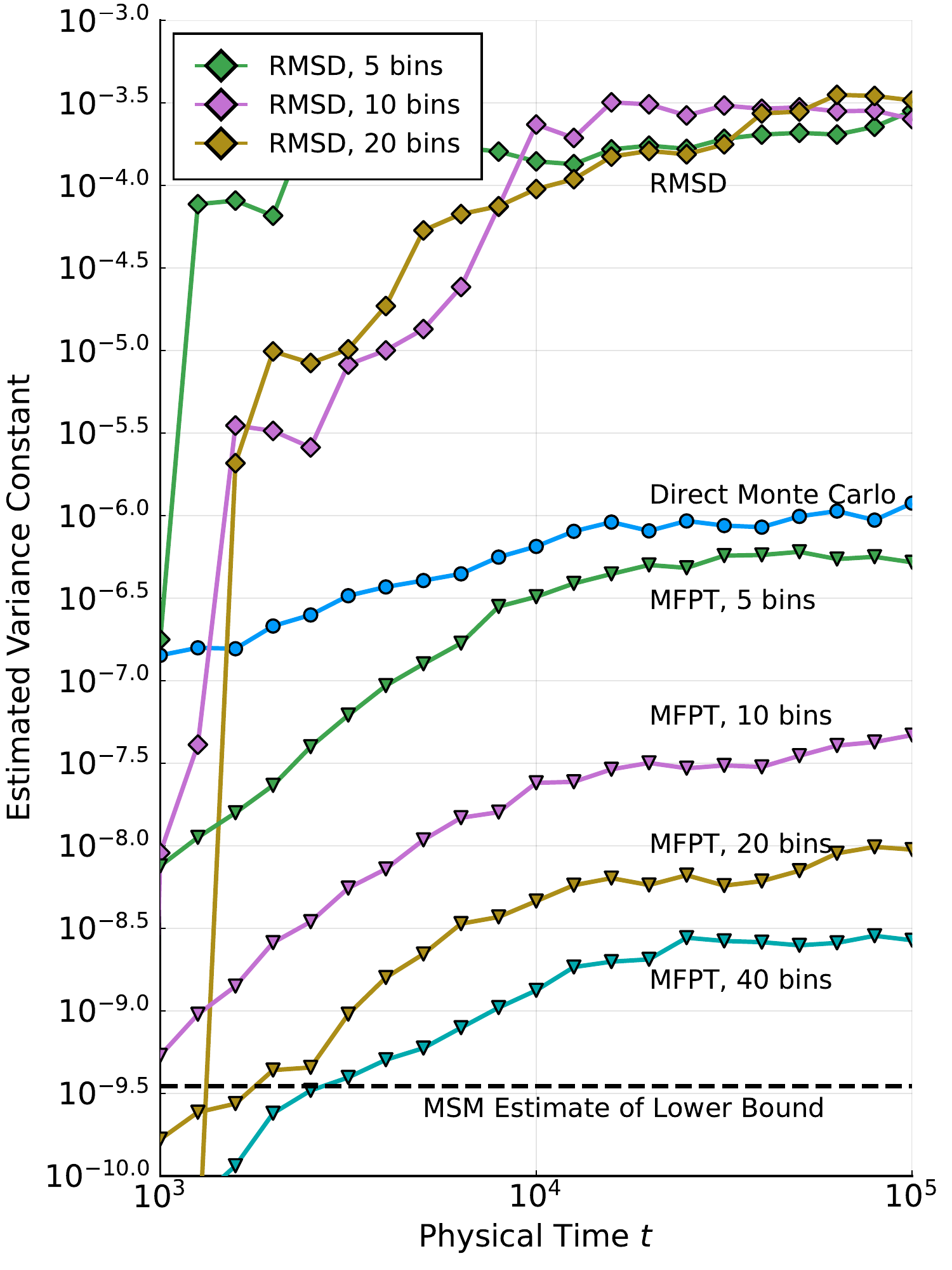}
    \caption{
    \emph{Comparison of binning strategies for MFPT estimation.}
    Estimates of the rescaled variance constant \eqref{e:rescaledvar1} are plotted for MFPT and RMSD binning strategies for the two-dimensional problem depicted in Fig.\ \ref{fig:msm2d}.  
    For this problem, the flux value is approximately $10^{-6}$.}
    \label{fig:varianceconst1}
\end{figure}

To evaluate the precision -- and hence, efficiency -- of the flux estimates from MFPT and RMSD binning, we define a compute-cost-independent variance constant:
\begin{align}\begin{split} \label{e:rescaledvar1}
    \textup{Variance Constant} = Nt \times \textup{Var}\left(\hat{J}_t\right).
    \end{split}
\end{align}
Here, the variance of the empirical 
WE flux has been scaled to account for total simulation cost $Nt$.
The variance constants for MFPT and RMSD binning are compared in Figure \ref{fig:varianceconst1}.
The figure also shows estimates of the optimal WE variance constant $(\int \pi v)^2$ from the fine-grained MSM model (see equation~\eqref{eq:opt_var_const}), as well as the variance constant for direct Monte Carlo,
without any splitting or merging.
We used $N = 10^4$ trajectories and averaged over $10^2$ independent runs to produce these variance constants.

Figure \ref{fig:varianceconst1} reveals a dramatic difference between MFPT and RMSD binning.
RMSD binning leads to worse-quality results than direct Monte Carlo simulation for this non-trivial model, for any number of bins.
Meanwhile, MFPT binning leads to a major improvement over direct Monte Carlo simulation,
particularly when the number of bins is large ($20$--$40$ bins).
With only $5$ bins, we see a modest gain with the MFPT binning strategy. 
Increasing the number of bins with MFPT binning systematically improves the output such that with $40$ bins, we are within an order of magnitude of the estimated optimal constant.
This is two and a half orders of magnitude of improvement over direct Monte Carlo. 

To encourage a fair comparison,
we have applied the optimal allocation rule \eqref{eq:opt_alloc_rule} to both MFPT and RMSD binning.
However, analogous tests (not pictured) show that RMSD binning performs even worse with uniform allocations, leading to higher variance constants. 
Our MFPT binning method here also outperformed related strategies in our earlier work~\cite{aristoff2020optimizing}.

\begin{figure}
    \centering
    \includegraphics[width=\columnwidth]{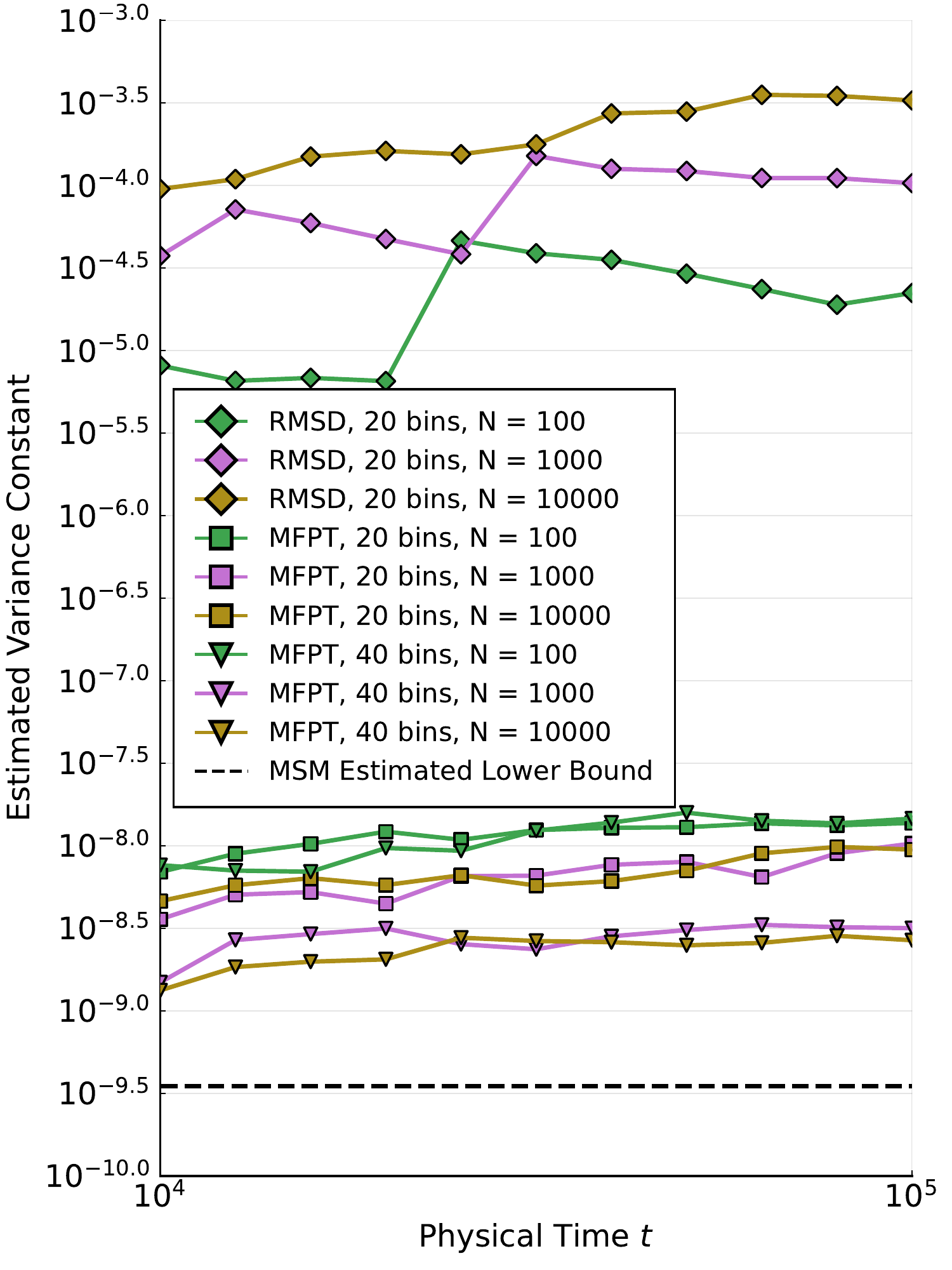}
    \caption{\emph{Variance constants for different numbers of trajectories}.}
    \label{fig:Nvarcomp}
\end{figure}

Lastly, the data in Figure \ref{fig:varianceconst1} was generated using $N = 10^4$ trajectories, which may exceed the computational resources available for problems of interest. 
Figure \ref{fig:Nvarcomp} presents comparisons for $10^2$, $10^3$, and $10^4$ trajectories, indicating that the variance constant typically has no more than an order of magnitude in variability. 
In particular, for the MFPT bins the results for $10^3$ trajectories are nearly indistinguishable from those obtained with $10^4$ trajectories.


\section{Conclusion} \label{sec:conclusion}

This work has attempted to assemble, expand upon, and make accessible to a chemical physics audience, recent mathematical results of fundamental importance to weighted ensemble (WE) simulation.
The mathematical theory exposes the fundamental capabilities of WE and guiding principles behind WE optimization.
It leads to the identification of optimal reaction coordinates for merging and splitting.
The theory also identifies the minimal possible variance for WE in the limit of many trajectories and time steps, and we demonstrate a simple binning strategy for approaching that optimum that works well in a two-dimensional model problem.
Testing these approaches in complex systems will be pursued in future work.

To illustrate the mathematical theory in a physically intuitive way,
explicit formulas were derived
for the optimal coordinates and 
optimal WE variance 
in the case of the Smoluchowski dynamics~\eqref{eq:Smoluchowski} in one dimension.
These formulas, presented for the first time here,
demonstrate that
the variance reduction from WE can be exponentially large in the size of the largest forward energy barrier from $A$ to $B$.

The mathematical theory is important because it enables principled variance reduction strategies for WE.
Here, we have proposed and tested one such variance reduction strategy, called MFPT binning.
After pilot runs that estimate $h$ and $v$ (or equivalently, the local MFPT to $B$ and its variance),
this strategy defines the bins to be intervals in the $h$ coordinate,
chosen in a way that optimizes the variance.
In a model problem, MFPT binning significantly outperforms a naive WE approach that constructs bins using the RMSD to $B$, and 
these results remain robust even 
for relatively small $N$.

\section*{Acknowledgements}

D. Aristoff and G. Simpson gratefully acknowledge support from the National Science Foundation via the awards DMS 2111277 and DMS 1818726. 
J. Copperman is a Damon Runyon Fellow supported by the Damon Runyon Cancer Research Foundation (DRQ-09-20).
R. J. Webber was supported by the Office of Naval Research through BRC award N00014-18-1-2363 and the National Science Foundation through FRG award 1952777, under the aegis of Joel A. Tropp. 
D. M. Zuckerman was supported by NIH grant GM115805. 
Computational resources were provided by Drexel’s University 673 Research Computing Facility.
The authors are grateful to 
Mats Johnson for preliminary 
numerical work related 
to the results in Section~\ref{sec:optimalWE}.

\section*{Appendix}
\appendix

\subsection{Computing other statistics using WE} \label{app:general}

WE can be used to compute other statistics, in addition to the MFPT from $A$ to $B$.
Namely, any integral $\int f(x) \pi(x) dx$
involving an observable $f$
can be estimated using $N$ WE trajectories.
In the long time limit, the trajectories satisfy
\begin{equation}
    \int f(x) \pi(x) dx
    = \lim_{t \rightarrow \infty} \frac{1}{t} \sum_{s=1}^t \sum_{i=1}^N w_{s\tau}^i f(\xi_{s\tau}^i),
\end{equation}
where $(\xi_t^i, w_t^i)_{1 \leq i \leq N}$
denotes the position and weight of trajectory $i$ at time $t$.
Therefore, as a practical estimator, we can terminate the simulations at a finite time $t$ and approximate
\begin{equation}
\label{eq:estimator}
    \int f(x) \pi(x) dx
    \approx \frac{1}{t-t_0} \sum_{s=t_0}^t \sum_{i=1}^N w_{s\tau}^i f(\xi_{s\tau}^i),
\end{equation}
where $t_0 > 0$ is a suitable burn-in time.
Here we have assumed $f$ is recorded only at $\tau$ intervals, although WE permits saving points more frequently.

The estimator \eqref{eq:estimator} is readily extended to history-dependent observables,
such as transition-path times or mechanistic pathways from $A$ to $B$.
This requires defining the trajectories as paths starting from $A$, which are recycled upon reaching $B$, and defining $\pi$ as a stationary measure on path space.
The recent mathematical literature~\cite{aristoff2016analysis,aristoff2020optimizing,aristoff2022ergodic,webber2020splitting} provides a set of error bounds and convergence guarantees for the estimator \eqref{eq:estimator}.

\subsection{Relationship between flux discrepancy and MFPT}
\label{sec:loc-mfpt}

To establish the relationship
\begin{equation}
    \lim_{t \to \infty} \left[\left\langle N_t \right\rangle_x - \left\langle N_t \right\rangle_{\pi} \right] 
    = \frac{\langle T_B \rangle_{\pi} - \langle T_B\rangle_x}{ \langle T_B\rangle_A},
\end{equation}
we first introduce the following terminology:
\begin{itemize}
    \item $T_B^x$ is the first passage time into $B$ for a dynamics started at $X_0 = x$.
    \item $T_B^\pi$ is the first passage time into $B$ for a dynamics started at $X_0 \sim \pi$, i.e., started at a point chosen randomly from the steady-state distribution.
    \item $N_t^A$ is the number of arrivals into $B$ by time $t$ for a dynamics started at $X_0 \sim \rho_A$.
\end{itemize}
Next we recall the \emph{renewal theorem} from classical probability~\cite{feller2008introduction}, which states that for any $s>0$,
\begin{equation}
    \lim_{t \rightarrow \infty} \left[ \langle N_t^A \rangle - \langle N_{t-s}^A \rangle \right] = \frac{s}{\langle T_B \rangle_A}.
\end{equation}
Lastly, we verify
\begin{align}
    & \lim_{t \to \infty} \left[\left\langle N_t \right\rangle_x - \left\langle N_t \right\rangle_{\pi} \right] \\
    &= \lim_{t \to \infty} \left[\left\langle N_{t-T_B^x}^A \right\rangle - \left\langle N_{t-T_B^\pi}^A \right\rangle \right] \\ 
    &= \lim_{t \rightarrow \infty}
    \left[\left\langle N_t^A \right\rangle
    - \left\langle N_{t-T_B^\pi}^A \right\rangle \right]
    -\lim_{t \to \infty} \left[\left\langle N_t^A \right\rangle 
    - \left\langle N_{t-T_B^x}^A \right\rangle \right] \\
    &= \frac{\langle T_B^{\pi} \rangle - \langle T_B^x\rangle}{\langle T_B\rangle_A},
\end{align}
where we have applied the renewal theorem after conditioning on $s = T_B^{\pi}$ and $s = T_B^x$.

\subsection{Derivation of optimal allocation rule}
\label{app:gen_var}

In this section we provide an intuitive derivation of the optimal allocation rule
\begin{equation}
\label{eq:optimal}
    \#\textup{ trajectories near }x \propto \pi(x)v(x).
\end{equation}

We introduce the following notation:
\begin{itemize}
    \item 
    $\textup{weight}_{\textup{bin}}$ and $\textup{alloc}_{\textup{bin}}$ indicate a bin weight and bin allocation.
    \item $\langle \rangle_{\textup{bin}}$ and $\textup{Var}_{\textup{bin}}$ denote 
    the mean and variance with respect to the weighted 
    trajectory distribution in a bin, so $\textup{Var}_{\textup{bin}}(f) = \langle f^2\rangle_{\textup{bin}} -\langle f\rangle_{\textup{bin}}^2$.
    \item Finally, $\langle \rangle_t$ 
    denotes the average over the WE ensemble up to time $t$.
\end{itemize}
Then the following variance formula~\cite{aristoff2022ergodic,webber2020splitting} holds in the limit as $t \rightarrow \infty$:
\begin{align}
& t \textup{Var}(\hat{J}_t) \sim \nonumber\\
&  \left\langle\sum_{\textup{bins}} \frac{\textup{weight}_{\textup{bin}}^2}{\textup{alloc}_{\textup{bin}}} \left[\textup{Var}_{\textup{bin}}( h)+ \textup{Var}_{\textup{bin}}(v)+\langle v\rangle_{\textup{bin}}^2\right]\right\rangle_t.\label{eq:WE_variance}
\end{align}

We consider a greedy minimization strategy~\cite{aristoff2020optimizing,aristoff2022ergodic} for minimizing the variance formula \eqref{eq:WE_variance}.
The variance terms $\textup{Var}_\textup{bin}(h)$ and 
$\textup{Var}_{\textup{bin}}(v)$ are minimized by choosing 
bins in which 
$h$ 
and $v$ values are nearly constant.
The other term, namely,
\begin{equation*}
    \frac{\textup{weight}_{\textup{bin}}^2}{\textup{alloc}_{\textup{bin}}} \langle v\rangle_{\textup{bin}}^2,
\end{equation*}
is minimized when the allocation is chosen using~\cite{aristoff2016analysis,aristoff2020optimizing,aristoff2022ergodic}
\begin{equation}\label{eq:opt_allocation}
   \textup{alloc}_{\textup{bin}}  \appropto \textup{weight}_{\textup{bin}}\times \langle v\rangle_{\textup{bin}},
\end{equation} 
which reduces to \eqref{eq:optimal} in the limit of many trajectories.

\subsection{Low temperature limit}
\label{app:low_temp}

Here we derive the optimal splitting and merging coordinates in the low-temperature limit $\beta \rightarrow \infty$.
First, we combine the expression for the steady state
\begin{equation}
    \ppi(x) 
    \propto e^{-\beta U(x)}\int_{\max\{x,a\}}^b \frac{e^{\beta U(y)} \mathop{dy}}{D(y)}
\end{equation}
with the expression for the flux variance function
\begin{equation}
    \tilde{v}(x) = \frac{\sqrt{2}e^{\beta U(x)}}
    {D(x)^{1 \slash 2} \langle \tilde{T}_B \rangle_A}
    \int_{-\infty}^x \frac{\mathop{dy}}{e^{\beta U(y)}},
\end{equation}
to obtain the optimal allocation rule
\begin{equation}
\label{eq:piv_1D}
    \ppi(x)\vv(x) \propto   \frac{1}{D(x)^{1 \slash 2}} \int_{\max\{x,a\}}^b \frac{e^{\beta U(y)} \mathop{dy}}{D(y)}
    \int_{-\infty}^x \frac{\mathop{dy}}{e^{\beta U(y)}}.
\end{equation}

Next, we assume the pair of points $(x_-, x_+)$ is the unique solution to the maximization
\begin{equation}
    \max_{a \leq y \leq b, \,x < y} (U(y) - U(x)),
\end{equation}
and $U^{\prime \prime}(x_+) < 0 < U^{\prime \prime}(x_-)$.
Using Laplace's method~\cite{Bender_Orszag_1999} on \eqref{eq:piv_1D} we obtain, for any $x_- < x < x_+$,
\begin{multline}
\label{eq:explicit_laplace}
    \frac{1}{D(x)^{1 \slash 2}} \int_{\max\{x,a\}}^b \frac{e^{\beta U(y)} \mathop{dy}}{D(y)}
    \int_{-\infty}^x \frac{\mathop{dy}}{e^{\beta U(y)}} \\
    \stackrel{\beta \rightarrow \infty}{\sim} \frac{C}{D(x)^{1 \slash 2}} \frac{ e^{\beta U(x_+)}}{\beta e^{\beta U(x_-)}},
\end{multline}
where the prefactor $C$ is explicitly
\begin{equation}
    C = \frac{2\pi}
    {D(x_+) U^{\prime \prime}(x_-)^{1 \slash 2} |U^{\prime \prime}(x_+)|^{1 \slash 2}}.
\end{equation}
For $x$ outside of the interval $[x_-, x_+]$, Laplace's method dictates that the optimal allocation $\pi(x) v(x)$ is exponentially smaller as $\beta \rightarrow \infty$, whence we recover the optimal allocation \eqref{eq:low_temp}.

From the above calculations,
we find that the optimal gain from using WE takes the form
\begin{equation}
    \textup{Gain over direct Monte Carlo} = \frac{QR}{S^2},
\end{equation}
where
\begin{align}
\label{eq:q}
    Q &= \int_{-\infty}^b 
    \frac{e^{\beta U(x)}}{D(x)}
    \int_{\max\{x,a\}}^b \frac{e^{\beta U(y)} \mathop{dy}}{D(y)}
    \Bigl(\int_{-\infty}^x
    \frac{\mathop{dy}}{e^{\beta U(y)}}\Bigr)^2 \mathop{dx}, \\
\label{eq:r}
    R &= \int_{-\infty}^b e^{-\beta U(x)}\int_{\max\{x,a\}}^b \frac{e^{\beta U(y)} \mathop{dy}}{D(y) } \mathop{dx}, \\
\label{eq:s}
    S &= \int_{-\infty}^b \frac{1}{D(x)^{1 \slash 2}} \int_{\max\{x,a\}}^b \frac{e^{\beta U(y)} \mathop{dy}}{D(y)}
    \int_{-\infty}^x \frac{\mathop{dy}}{e^{\beta U(y)}} \mathop{dx}.
\end{align}
Applying the Laplace principle to \eqref{eq:q}-\eqref{eq:s} 
verifies the expression
\begin{multline}
    \textup{Gain over direct Monte Carlo} \\
    \sim \frac{\pi/\beta}{\left(\int_{x_-}^{x_+} \sqrt{D(x_+) \slash D(x)}\,dx\right)^2}\frac{e^{\beta (U(x_+)-U(x_-))}}{\sqrt{|U''(x_-)U''(x_+)|}}.
\end{multline}

\subsection{Details of numerical simulations} \label{app:sim_details}

In the 2D problem,  the landscape is given by the expression
\begin{equation}
    U(x, y) = U_1(x, y) + U_2(x, y) + 0.5 U_3(x, y),
    \label{utwod}
\end{equation}
where
\begin{subequations}
\begin{align}
    \begin{split}
    \log U_1(x,y) =&  -c_1 (x-0.25)^2 - c_1(y-0.75)^2 \\
    & -2c_2(x-0.25)(y-0.75)
    \end{split} \\
    \log U_2(x, y) =& -c_3 x^2 (1-x)^2 y^2(1-y)^2 \\
    \log U_3(x, y) =& -c_4x^2 - c_4 y^2 + 2c_5x y
\end{align}
\end{subequations}
and the constants are given by
\begin{equation}
    (c_1, c_2, c_3, c_4, c_5) =  (50.5, 49.5, 10^5, 51, 49).
\end{equation}
The diffusivity is $D=1$, and the inverse temperature is $\beta = 30$.

To simulate the Smoluchowski dynamics, we apply Euler-Maruyama integration with an integration step of size $0.001$. Each evolution step is composed of ten such steps, so that $\tau = 0.01$. 
We constrain the trajectories to reside in $[0,1]^2$.
In the case that a trajectory attempts to exit this region, it is projected back into the box with $x=\max(\min (x,1),0)$ and $y=\max(\min (y,1),0)$.

The functions $\pi$, $h$, and $v$ are approximated using an MSM model generated from ``microbins''~\cite{copperman2020accelerated,aristoff2020optimizing} with Voronoi cell centers that are spaced $\Delta x = \Delta y = 0.02$ units apart on a regular Cartesian grid.
This spacing results in a total of $49 \times 49 = 2401$ microbins, which are the small rectangles visible in Figure~\ref{fig:msm2d}.


\bibliography{aipsamp}

\end{document}